\documentstyle[12pt,epsfig]{article}                                
\begin{document}

\titlepage                  

\begin{center}
\begin{Large}
\begin{bf}

Two different analysis on derivation of  PYTHAGORAS Theorem  (569 -479 B.C): Discrete continuum states

\vspace{1.0cm}

\end{bf}
\end{Large}

 Biswanath Rath 

\end{center}

\vspace{0.1cm}

\begin{it}
 Department of Physics,
 Maharaja Sriram Chandra Bhanj Deo University,
 Takatpur, Baripada -757003, Odisha, India.
e.mail:biswanathrath10@gmail.com

\vspace{0.1cm}

\end{it}

$\bf{Abstract:}$
We prese3nt two different ways to derive PYTHOGORAS theorem without assuming 
right angle concept. In each case we generate a model quantum well and notice that energy levels are discrete but continuous in nature. We also present wave function nature in each case.
  
\vspace{1.0cm}

\begin{bf}

\hspace{0.10cm}


\end{bf}


\vspace{0.1cm}

\hspace{0.5cm} \noindent\rule{3.0in}{0.4pt}

Correspondence: biswanathrath10@gmail.com 

\begin{bf}
1.Introduction
\end{bf}

Nearly 4000 years ago, a famous Greek mathematician cum philosopher named " Pythogoras" proposed that if sum of  squares of two sides of a triangle becomes squares of the third side, then the triangle must be "right angle triangle".
 Mathematically 
\begin{equation}
a^{2}+b^{2}=c^{2}
\end{equation}
where $a,b$ and $c$ are the sides of the triangle.

\vspace{1.0cm}
\begin{table}
\begin{center}
\begin{tabular}{c  } \\
\includegraphics[width= 0.5\textwidth]{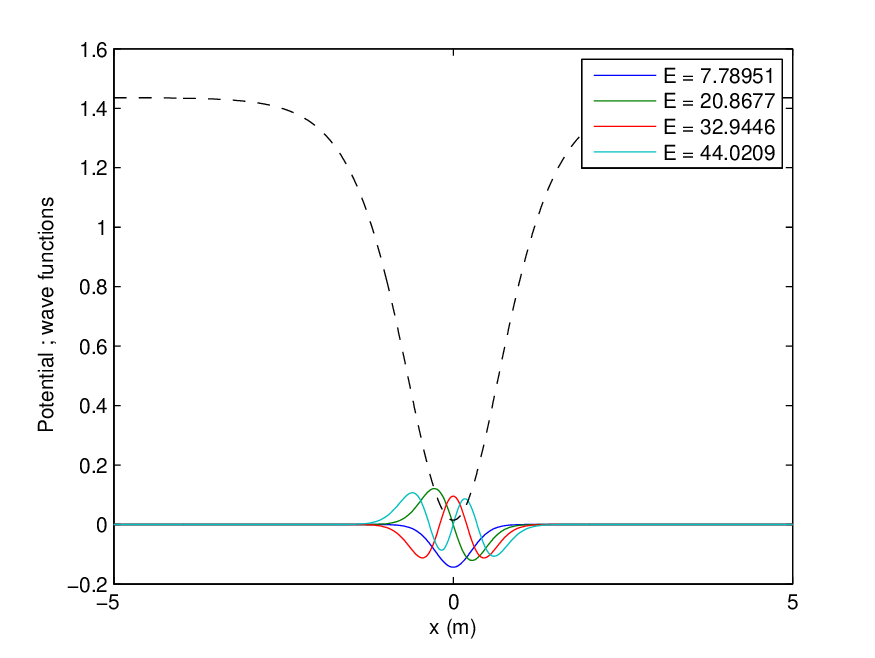} \\
\hspace{0.1cm}Fig.1. Any triangle    \\
\end{tabular}
\end{center}
\end{table}

Derivation of this relation is still interesting. In fact practical application of this theorem to quantum physics  is still lacking interest. In thiis paper we propose two different ways to derive the theorem, In each case a quantum well has been proposed to study nature of quantum states as follows

\begin{bf}
2A. Model derivation using $\tanh(x)$ and $sech(x)$
\end{bf}

Let us define the following 
\begin{equation}
\frac{a}{c}= \frac{1}{\cosh(x)} \rightarrow  a= c sech(x)
\end{equation}
\begin{equation}
\frac{b}{c}= \tanh(x)
\end{equation}
Hence it is to show that 
\begin{equation}
b=a \sinh(x)
\end{equation}
Hence using the relation 

\begin{equation}
a^{2}+b^{2}=a^{2} (\cosh(x))^{2} = [ c^{2} (sech(x))^{2}] [\cosh(x)]^{2}=c^{2}
\end{equation}

\begin{equation}
a^{2}+b^{2} = c^{2}
\end{equation}

Hence the theorem is proved.

\begin{bf}
 2,B- Model potential and Spectral nature
\end{bf}

 Here we construct a model potential as 

\begin{equation}
U(x) = \frac{a^{2}}{c^{2}} + \lambda \frac{b^{2}}{c^{2}}
\end{equation}

whose explict form is 
\begin{equation}
U(x) =  (sech(x))^{2} + \lambda (\tanh(x))^{2}
\end{equation}

The corresponding Hamiltonian for $\lambda=100$ is 

\begin{equation}
H = p^{2} + U(x)
\end{equation}

In this case energy levels are discrete and reflect continuum behaviour as seen below.

\begin{table}
\begin{center}
\begin{tabular}{c c } \\
\includegraphics[width= 0.5\textwidth]{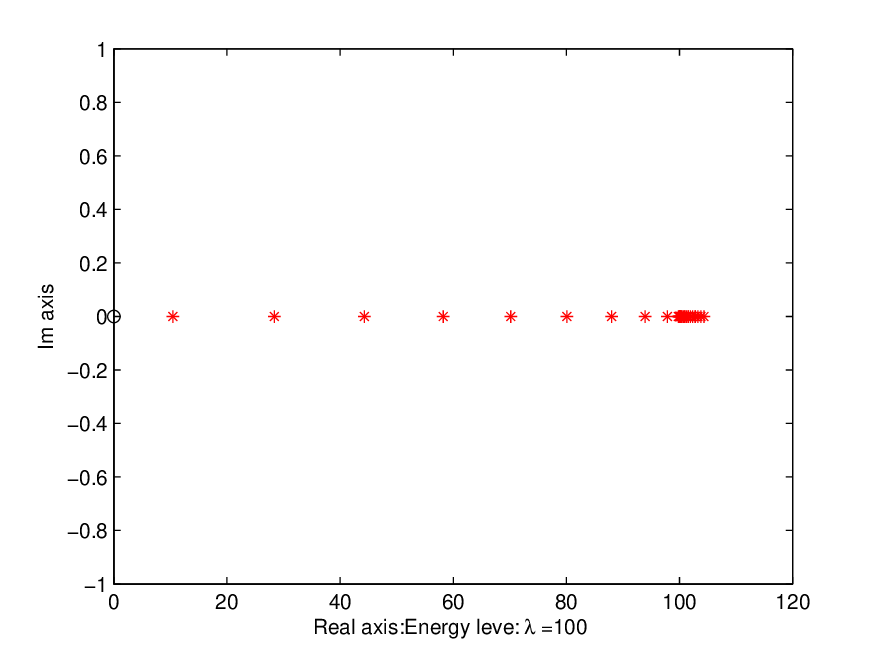} &  
\includegraphics[width= 0.5\textwidth]{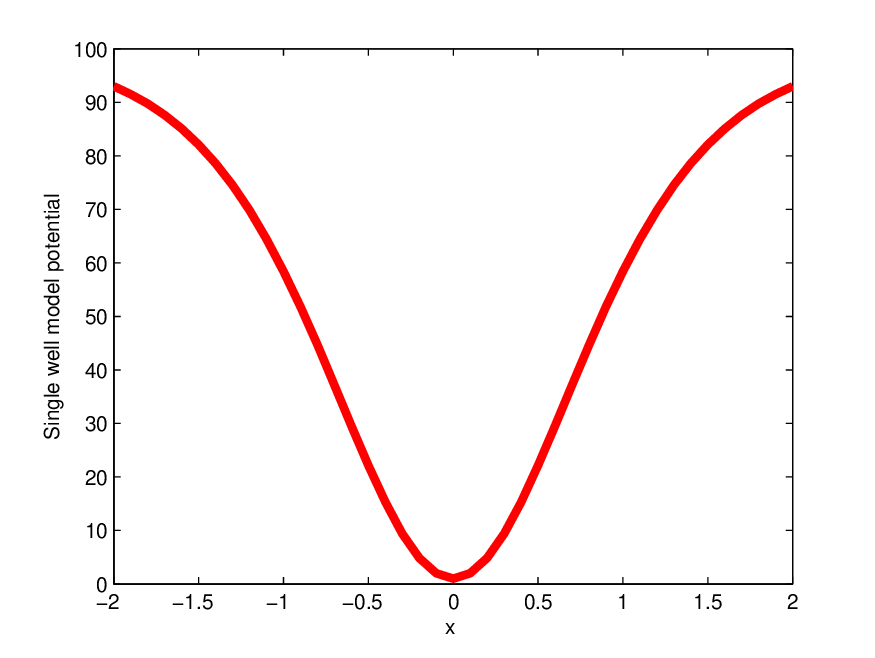}   \\
\includegraphics[width= 0.5\textwidth]{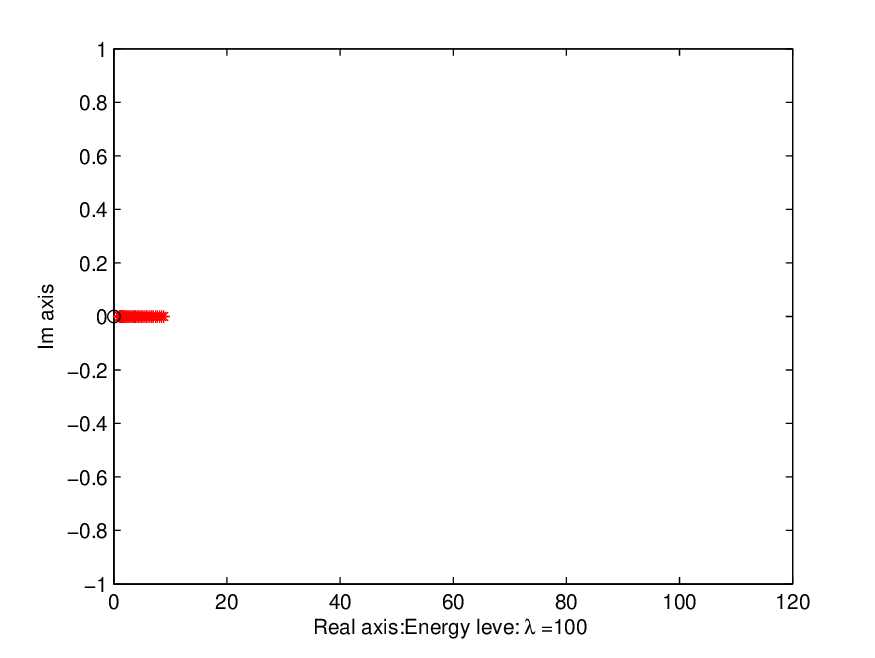} &  
\includegraphics[width= 0.5\textwidth]{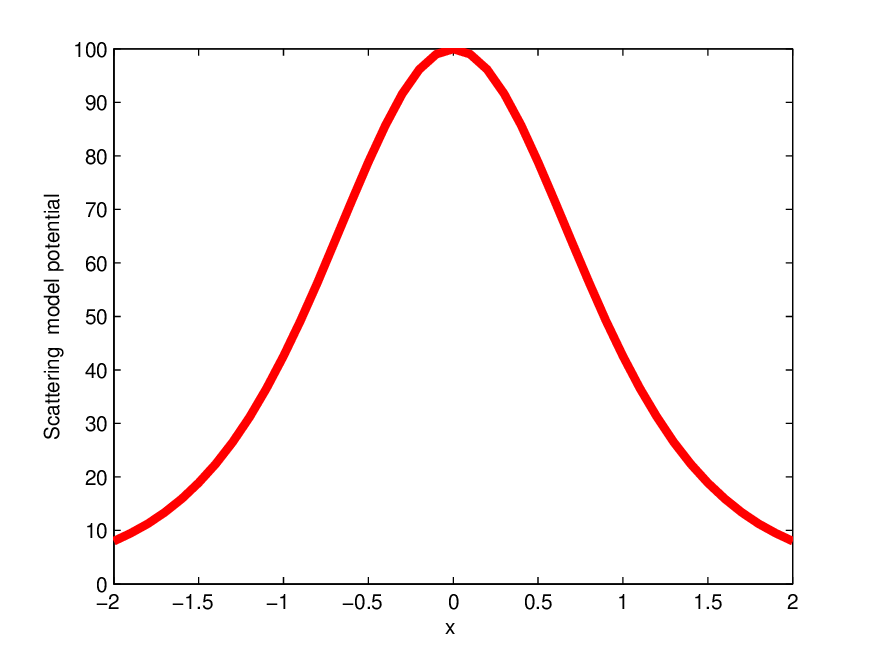}   \\
\hspace{0.1cm}Fig.2. Wave function and continuum states   \\
\end{tabular}
\end{center}
\end{table}

The first four eigenstates energy are given below 

\begin{table}
Table-1.First four  energy levels 
\begin{center}
\begin{tabular}{c c c  } \\
 n & Energy & Energy (scattering model) \\ \hline
0 & 10.4624 & 1.0080\\
1 & 28.3872 & 1.0210 \\
2 & 44.3121 & 1.0494 \\
3 & 50.2370 & 1.0840\\ \hline
\end{tabular}
\end{center}
\end{table}

\begin{bf}
3A. Second derivation of Pythagoras theorem using $\sinh(x)$ and $\cosh(x)$
\end{bf}

 Here, we consider another model derication of Pythagoras theorem using 

\begin{equation}
a^{2}+ b^{2}= c^{2}
\end{equation}
Mathematically consider that 

\begin{equation}
\frac{b^{2}}{c^{2}}= \frac{[\cosh(x)]^{2}}{\cosh(2x)}
\end{equation}
and 
\begin{equation}
\frac{a^{2}}{c^{2}}= \frac{[\sinh(x)]^{2}}{\cosh(2x)}
\end{equation}

Now adding we get
\begin{equation}
\frac{b^{2}}{c^{2}}+\frac{a^{2}}{c^{2}}= \frac{\cosh^{2} x}{\cosh(2x)}+ \frac{\sinh^{2} x}{\cosh(2x)} = 1
\end{equation}
Hence 
\begin{equation}
a^{2}+b^{2}=c^{2}
\end{equation}

\begin{bf}
3B. Potential model  and Bound states
\end{bf}

Here, we choose the potential as 

\begin{equation}
V(x) = \frac{a^{2}}{c^{2}} + \lambda \frac{b^{2}}{c^{2}}
\end{equation}

Here the corresponding Hamiltonian is written as 

\begin{equation}
H= p^{2} + V(x)
\end{equation}
 in its explict form 

\begin{equation}
H=p^{2}+ \frac{[ \cosh(x)]^{2}}{\cosh(2x)} + \lambda  \frac{[\sinh(x)]^{2}}{\cosh(2x)}
\end{equation}

For $\lambda=100$ we  present potential , wave function nature and duiscrete energy levels nature. below 

\begin{table}
\begin{center}
\begin{tabular}{c c } \\
\includegraphics[width= 0.5\textwidth]{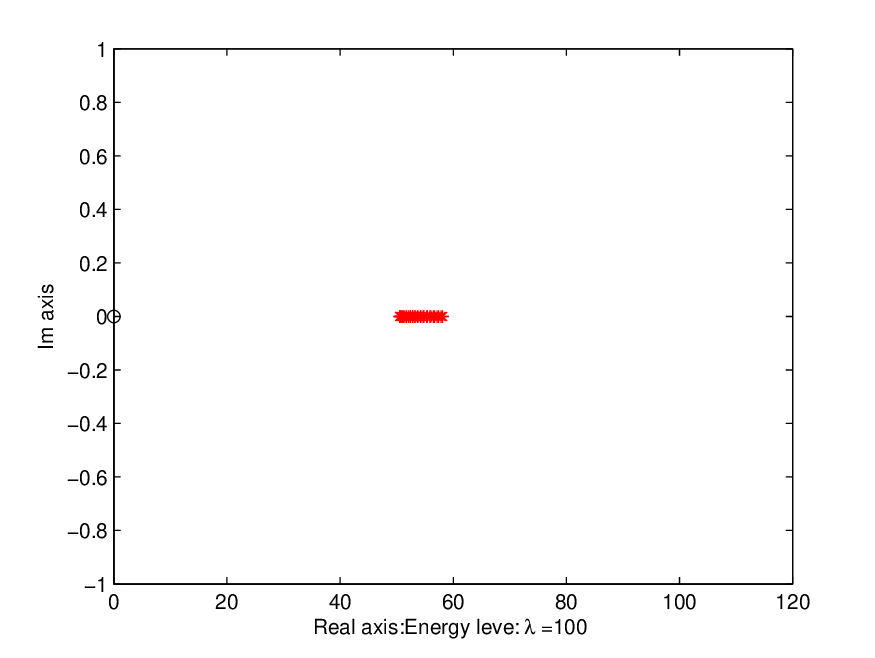} &  
\includegraphics[width= 0.5\textwidth]{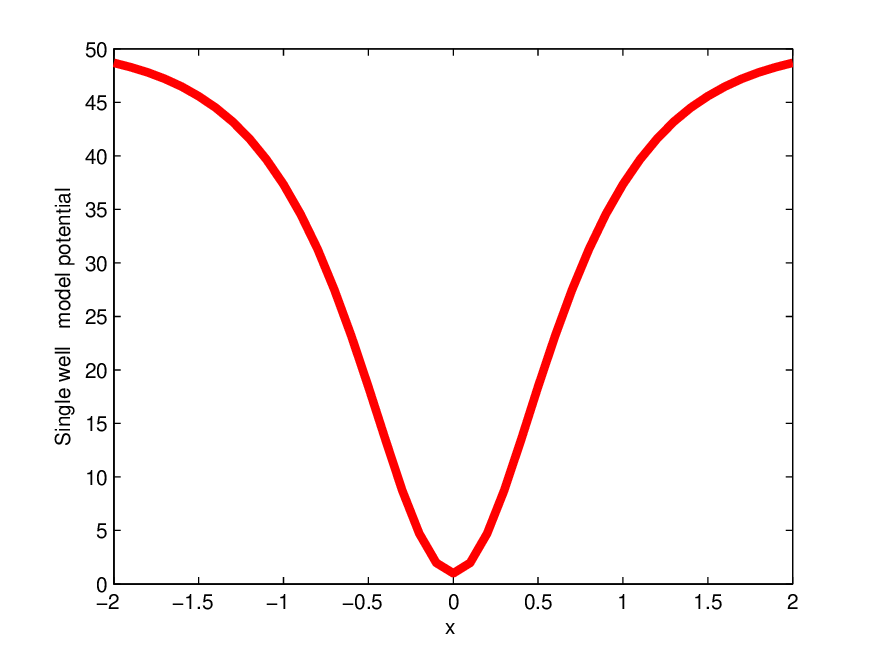}   \\
\includegraphics[width= 0.5\textwidth]{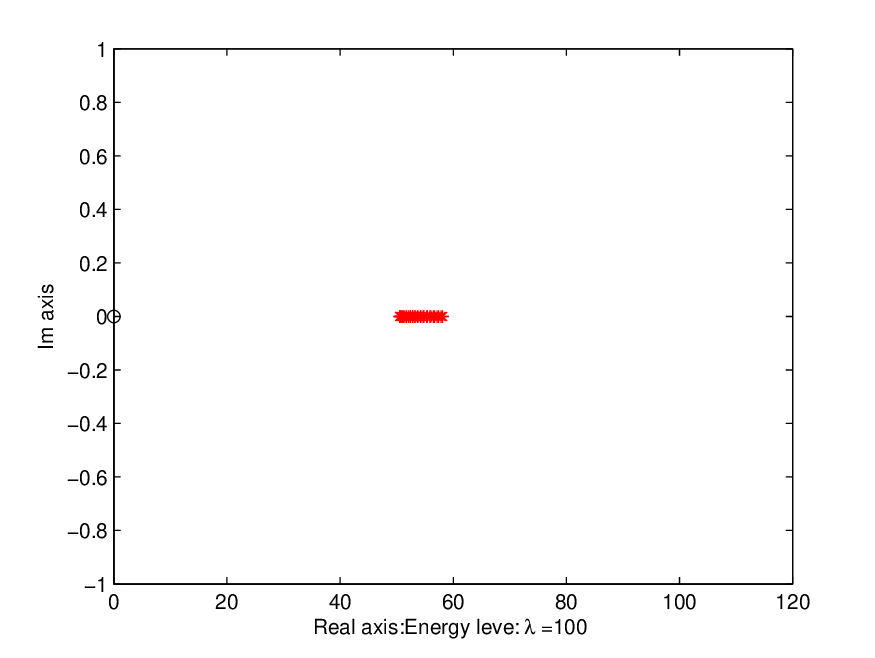} &  
\includegraphics[width= 0.5\textwidth]{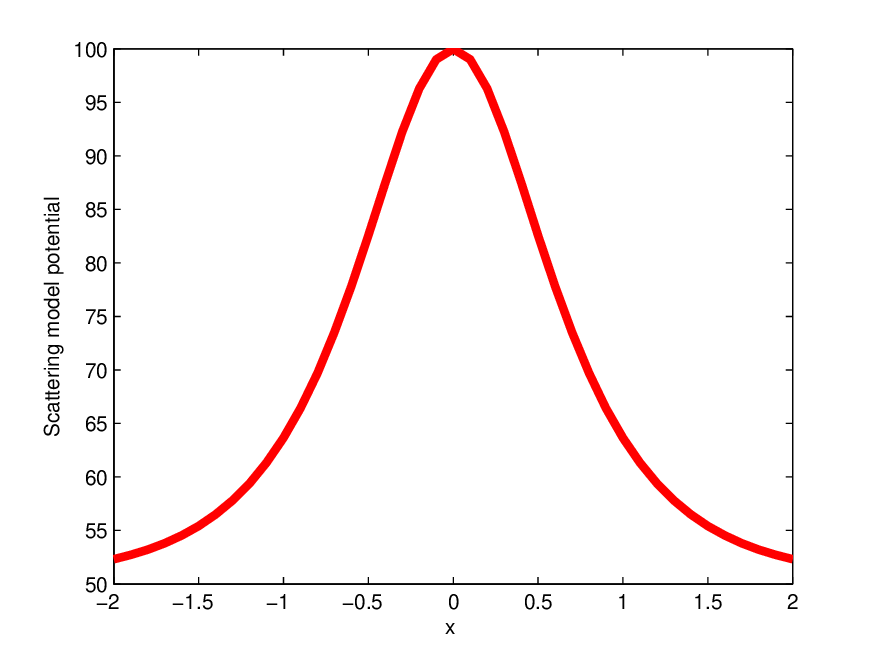}   \\
\hspace{0.1cm}Fig.2. Wave function and continuum states   \\
\end{tabular}
\end{center}
\end{table}

 The first four eigenstates energy are given below 

\begin{table}
Table-2. First four energy levels \\
\begin{center}
\begin{tabular}{c c c } \\ \hline 
 n & Energy & Energy (Scattering model) \\  \hline
0 & 9.8265 & 50.5113 \\
1 & 25.2187 & 50.5121 \\
2 & 29.8358 & 50.5454 \\
3 & 36.3890 & 50.5486 \\  \hline
\end{tabular} 
\end{center}
\end{table}

\pagebreak

\begin{bf}
4. Method of calculation
\end{bf}

Here we solve the eigenvalue relation
\begin{equation}
H \Psi = E\Psi
\end{equation}

where 
\begin{equation}
|\Psi>= \sum A_{m} |m>
\end{equation}
Here $m>$ satisfies the eigenvalue relation

\begin{equation}
[p^{2}+x^{2}]|m> =(2m+1)|m>
\end{equation}

\begin{bf}
5.Conclusion
\end{bf}
We have presented two different ways to derive the PYTHAGORAS theorem without assuming the concept of right angle triangle.In thhis derivation, we use hyperbolic functions. Further model Hamiltonian reported here shows discrete energy levels on quantum well and inverted scattering model potential. In fact continuum levels are well demarcated in lower and higher energy levels. However in the case of well continuum energy lavels are seen in higher quantum states.
\pagebreak

\begin{bf}
Author's contribution:
\end{bf}

B.Rath: formulation,computation,writing,finalization.

\begin{bf}
Conflict of interest
\end{bf}

Author declares there is no conflict of interest.

\begin{bf}
DATA AVAILABILITY
\end{bf}

No additional data is required . All the datas  included in this paper are 
sufficient.

\begin{bf}
Declaration
\end{bf}

Present paper is a modified version of arxiv paper.

\end{document}